\documentclass[aps,twocolumn,prl,superscriptaddress]{revtex4} %

\usepackage{natbib}
\usepackage{graphicx}
\usepackage{subfigure}
\usepackage{setspace}
\usepackage{umoline}
\usepackage{color}
\usepackage{verbatim}

\usepackage{amsbsy}

\newcommand{\ket}[1]{\left | #1 \right \rangle}

\begin{document}

\title{Quantum interference and manipulation of entanglement in silicon wire waveguide quantum circuits}
\author{Damien Bonneau}
\affiliation{Centre for Quantum Photonics, H. H. Wills Physics Laboratory \& Department of Electrical and Electronic Engineering, University of Bristol, Merchant Venturers Building, Woodland Road, Bristol, BS8 1UB, UK}
\author{Erman Engin}
\affiliation{Centre for Quantum Photonics, H. H. Wills Physics Laboratory \& Department of Electrical and Electronic Engineering, University of Bristol, Merchant Venturers Building, Woodland Road, Bristol, BS8 1UB, UK}
\author{Kazuya Ohira}
\affiliation{Corporate Research \& Development Center, Toshiba Corporation, 1, Komukai Toshiba-cho, Saiwai-ku, Kawasaki 212-8582, Japan}
\author{Nob Suzuki}
\affiliation{Corporate Research \& Development Center, Toshiba Corporation, 1, Komukai Toshiba-cho, Saiwai-ku, Kawasaki 212-8582, Japan}
\author{Haruhiko Yoshida}
\affiliation{Corporate Research \& Development Center, Toshiba Corporation, 1, Komukai Toshiba-cho, Saiwai-ku, Kawasaki 212-8582, Japan}
\author{Norio Iizuka}
\affiliation{Corporate Research \& Development Center, Toshiba Corporation, 1, Komukai Toshiba-cho, Saiwai-ku, Kawasaki 212-8582, Japan}
\author{Mizunori Ezaki}
\affiliation{Corporate Research \& Development Center, Toshiba Corporation, 1, Komukai Toshiba-cho, Saiwai-ku, Kawasaki 212-8582, Japan}

\author{Chandra M. Natarajan}
\affiliation{Scottish Universities Physics Alliance and School of Engineering and Physical Sciences, Heriot-Watt University, Edinburgh, EH14 4AS, United Kingdom}
\author{Michael G. Tanner}
\affiliation{Scottish Universities Physics Alliance and School of Engineering and Physical Sciences, Heriot-Watt University, Edinburgh, EH14 4AS, United Kingdom}
\author{Robert H. Hadfield}
\affiliation{Scottish Universities Physics Alliance and School of Engineering and Physical Sciences, Heriot-Watt University, Edinburgh, EH14 4AS, United Kingdom}
\author{Sanders N. Dorenbos}
\affiliation{Kavli Institute of Nanoscience, TU Delft, 2628CJ Delft, The Netherlands}
\author{Val Zwiller}
\affiliation{Kavli Institute of Nanoscience, TU Delft, 2628CJ Delft, The Netherlands}

\author{Jeremy L. O'Brien}
\affiliation{Centre for Quantum Photonics, H. H. Wills Physics Laboratory \& Department of Electrical and Electronic Engineering, University of Bristol, Merchant Venturers Building, Woodland Road, Bristol, BS8 1UB, UK}
\author{Mark G. Thompson}
\email{Mark.Thompson@Bristol.ac.uk}
\affiliation{Centre for Quantum Photonics, H. H. Wills Physics Laboratory \& Department of Electrical and Electronic Engineering, University of Bristol, Merchant Venturers Building, Woodland Road, Bristol, BS8 1UB, UK}

\date{\today}

\begin{abstract}

Integrated quantum photonic waveguide circuits are a promising approach to realizing future photonic quantum technologies. Here, we present an integrated photonic quantum technology platform utilising the silicon-on-insulator material system, where quantum interference and the manipulation of quantum states of light are demonstrated in components orders of magnitude smaller than previous implementations.  Two-photon quantum interference is presented in a multi-mode interference coupler, and manipulation of entanglement is demonstrated in a Mach-Zehnder interferometer, opening the way to an all-silicon photonic quantum technology platform.

\end{abstract}

\maketitle

\section*{Introduction}
Quantum information technologies offer completely new approaches to encoding, processing and transmitting information. By harnessing the properties of quantum mechanics, such as superposition and entanglement, it has been shown possible to realise fundamentally new modes of computation \cite{NielsenChuang,Deutsch-85}, simulation \cite{Feynman1982,Lloyd_QS96} and communication \cite{Gisinqc-np-2007}, as well as enhanced measurements and sensing \cite{gv-qm-sc-2004}. Of the many prospective physical systems in which to encode quantum information, photons are a particularly promising approach due to their properties of low noise, easy  manipulation and low transmission losses. To date, quantum photonics integrated circuits have been realised in low index contrast waveguide material systems, such as silica  \cite{Mataloni2011IntegratedPolarization,Shadbolt-np-2011} and silicon-oxy-nitride  \cite{ap-sc-qw2010}.  Such technologies offer benefits in terms of low propagation losses, but their associated large bend radii limits the scalability and usefulness of this technology.

Here we present silicon quantum photonic waveguide circuits utilising the silicon-on-insulator material system, where quantum interference and the manipulation of quantum states of light were demonstrated in components orders of magnitude smaller than previous implementations.  Quantum interference of indistinguishable photons was realised in multi-mode interference couplers, and manipulation of multiphoton entanglement was demonstrated in a Mach-Zehnder interferometer.

\section*{Integrated quantum photonics}
Traditionally, photonic quantum information experiments have been implemented using bulk optical elements, with photons propagating in free space. Although many proof-of-principle experiments have been reported, such an approach rapidly becomes impractical as the complexity of the quantum optical circuits increases, thereby making them inherently unscalable and confining them to research laboratory optical tables. In addition, sub-wavelength stability is critical for reliable operation of many quantum circuits due to the necessity of often complex networks of nested interferometers. However, developments over the past few years have overcome these bottlenecks through implementation of integrated quantum circuits, allowing quantum information science experiments to be realised that are inherently stable and orders of magnitude smaller than their equivalent bulk optic implementations.Integrated quantum photonic circuits have been demonstrated in a host of different material systems, including silica-on-silicon \cite{Po-Science-320-646,Shadbolt-np-2011}, direct write silica \cite{iw-OE-17-13519,Mataloni2011IntegratedPolarization}, silicon oxy-nitride \cite{ap-sc-qw2010}, lithium niobate \cite{bd-arXiv-LN} and gallium nitride \cite{ZY-GaNQI-2011}, all realized in low index-contrast waveguide structures. Two-photon quantum interference has been reported in both silica and gallium nitride \cite{Po-Science-320-646,ZY-GaNQI-2011}, whilst manipulation of quantum states of light has been demonstrated using silica-on-silicon \cite{Matt-NP-3-346,Shadbolt-np-2011}  and lithium niobate \cite{bd-arXiv-LN}; the later being also a promising material for photon pair generation \cite{NiceAboussouan2010}. The silicon-on-insulator wire waveguide technologies offer a further level of control, stability and miniaturisation, and also allow routes for on-chip generation and detection of photons. The high refractive index contrast provides compactness unrivalled by any other photonic material system, allowing a dramatic reduction in the footprint of quantum circuits and the integration of complex circuits on a single chip. Thermo-optic heaters, p-n junction modulators and induced $\chi^2$ nonlinearities \cite{Nature2006SiliconKhi2} allow for dynamically reconfigurable circuits with the possibility of fast phase control. The high third-order nonlinearity of silicon enables on-chip generation of quantum states of light via spontaneous four wave mixing \cite{sh-pg-oe-2006}, thus allowing integration of single photon sources and waveguide circuits on the same chip. Recently, high efficiency on-chip single photon detectors integrated with silicon waveguides have been reported \cite{Pernice2011High}. Demonstration of quantum interference and manipulation of quantum states of light in a silicon waveguide circuit is the next critical step to realizing photonic quantum technologies in silicon.In this work we report quantum interference and phase manipulation of one and two photon states using thermo-optic phase shifters and multimode interference devices on a silicon chip, opening up the way to a CMOS compatible silicon-based photonic quantum technology platform, where sources, detectors and circuits can all be realized monolithically on the same integrated chip for applications in communication, computing and metrology.

\section*{Experimental detail}
\begin{figure}
   \includegraphics[width=\columnwidth]{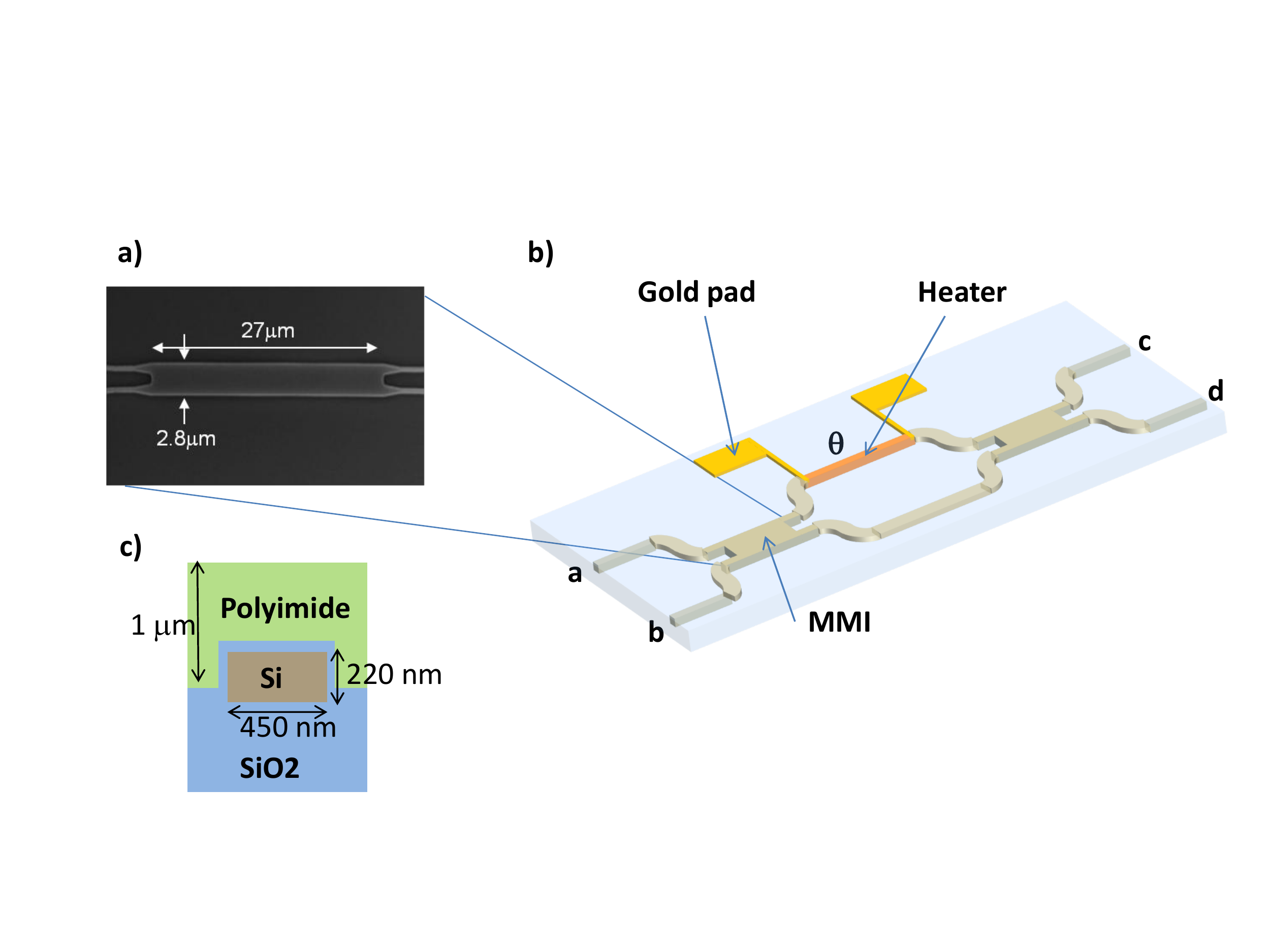}
    \caption{
    \label{Schemes}
   \textbf{MMI and MZI}
	a) SEM image of a multimode interference device, b) Schematic diagram of waveguide circuit with voltage controlled phase shifter, c) Illustration of the cross-section of the single-mode waveguide.
	}
\end{figure}

The silicon integrated waveguide circuits were fabricated from a silicon-on-insulator (SOI) substrate with a silicon thickness of 220nm.  The single-mode optical waveguides had a width of 450 nm, and a top cladding of silicon dioxide. Input and output coupling to lensed fiber was achieved using spot-size converters (SSC) comprising a 300 $\mu$m-long inverse-taper with a 200 nm tip width and a 4x4 $\mu$$m^2$ polymer waveguide. The silicon waveguiding structures were defined by 248 nm lithography and formed by dry etch processing. Typical propagation losses of 3.5 dB/cm and SSC losses of 2 dB/facet where observed, leading to device losses in the range of $6-10$ dB.
To realise the beam-splitter-like operation required for quantum interference, a 2x2 multi-mode interference coupler was designed (using FDTD simulations) and implement with dimensions of 2.8 $\mu$m x 27 $\mu$m and input tapers of length 3 $\mu$m and width 1 $\mu$m (Fig.~\ref{Schemes}(a)). Mach-Zehnder interferometers were formed from two multimode interference couplers and a 200 $\mu$m-long thermo-optic phase shifter, shown in Fig.~\ref{Schemes}(b). By varying the voltage across the phase shifter it was possible to tune the internal phase delay of the Mach-Zehnder interferometer. Pairs of indistinguishable single photons at a wavelength of 1550~nm were generated from a type-I spontaneous parametric down conversion source (SPDC). A bismuth borate BiB$_3$O$_6$ (BiBO) nonlinear crystal was pumped by a 775~nm, 150~fs pulsed Ti-Sapphire laser, and degenerate photon pairs were collected from two diametrically opposite points on the SPDC cone using single mode fibers,
and coupled into the chip using lensed fibers. One arm of the collection system was on a motorized delay stage to provide a tunable delay between the photon pair. The 1550nm photons were coupled out of the chip using single-mode lensed fiber and detected using two superconducting single photon detectors (SSPD), having system detector efficiencies of 5\% and 15\%\cite{Ha-OptE-13-10846,dsn-APL-93-131101}.
 Coincidence detection between the two output ports of the device was performed using a custom made (FPGA based) counting logic with a 5 ns coincidence window.

\section*{Quantum interference in a 2x2 multimode interference coupler}

Quantum interference lies at the heart of linear quantum photonics and is fundamental to the implementation of any photonic quantum technology, describing at it's most basic level the interaction of indistinguishable photons incident at a beamsplitter. In the case of two identical photons, each entering separate ports of a 50:50 beamsplitter device, the amplitude probabilities of both photons exiting different ports are zero $-$ giving a superposition state of either both photons exiting one output port, or both photons exiting the other output port. This quantum interference occurs for any physical implementation of a two-port beamsplitter device, and for integrated waveguide circuit two typical implementations are the directional coupler and multimode interference coupler. 

\begin{figure}
   \includegraphics[width=\columnwidth]{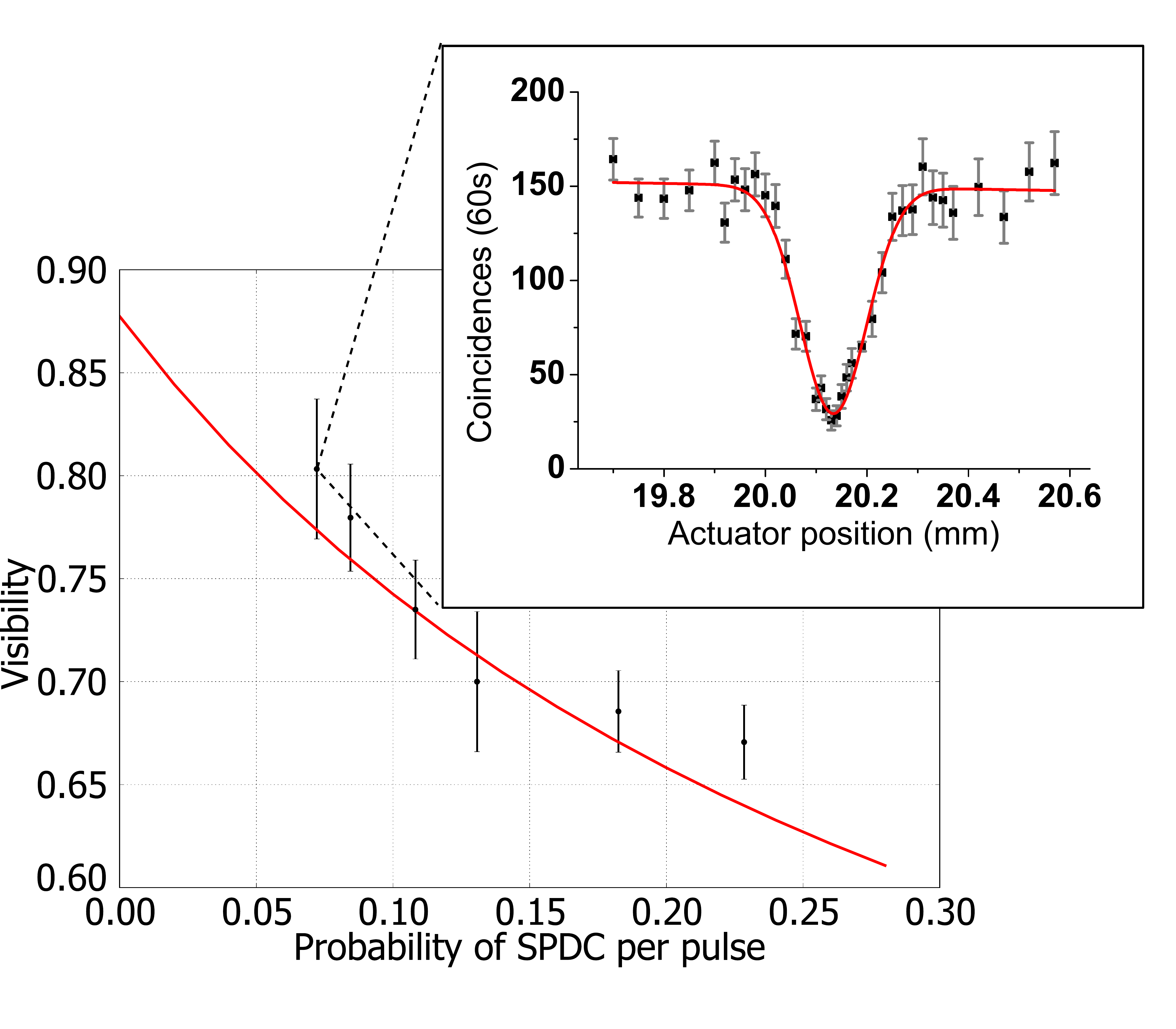}
    \caption{
    \label{VisibilityvsProbaSPDC}
   \textbf{Visibility as a function of the probability of getting a photon pair} Visibility of the Hong-Ou-Mandel experiment is plotted as a function of the probability per pulse of generating a photon pair. Solid line is a theoretical fit using a model which accounts for multi-photon events and losses before and after the MMI. Inset shows the two-photon interference plot for the highest visibility measurement.
	}
\end{figure}

Multimode interference (MMI) devices are based on the self-imaging principle, and compared with directional couplers they have improved tolerances to fabrication, a wider spectral bandwidth of operation, and can be extended to realise NxN multiport devices with several input and output ports \cite{Pe-JoLT-13-615}. To demonstrate quantum operation of the silicon-on-insulator 2x2 MMI coupler, indistinguishable photon pairs were launched into each of the two input waveguides. At the two output ports, the two-photon coincidence counts were monitored. By varying the temporal delay between the two input photons, the indistinguishability of the input state could be controlled and the Hong-Ou-Mandel (HOM) dip \cite{Hon-PRL-59-2044} could be observed (see inset of Fig.~\ref{VisibilityvsProbaSPDC}). A maximum dip visibility of $~80\%\pm3\%$ was observed, demonstrating quantum interference. 

A limited visibility of 80\% was achieved in part due to multi-photon events in the source, in part due to intrinsic losses within the MMI device, and also due to residual distinguishability of the generated photon pairs. The indistinguishability of the SPDC generated photon pairs was characterised by performing a HOM experiment with a bulk-optic beamsplitter, giving a maximum visibility of $95.5\%\pm1.5\%$ for low pump power. Due to the chip insertion losses (a total of 9 dB including input/output coupling) and detector efficiencies (5\% and 15\%), it was necessary to pump the SPDC source in a regime where high count rates could be achieved, which gave rise to multi-photon events that reduced the visibility of the dip. In order to quantify the effect of these multiphoton events, the visibility at different pump powers was measured (see Fig.~\ref{VisibilityvsProbaSPDC}), clearly demonstrating improved visibility for reduced photon-pairs per pulse. By fitting to a model which accounts for multi-pair generation and system losses (see Appendix), an extrapolate nominal visibility of $88\%\pm3\%$ at low pump power was demonstrated.

The 7.5\% discrepancy between the measured beamsplitter visibility of $95.5\%\pm1.5\%$ and the nominal MMI visibility of $~88\%\pm3\%$ can be accounted for by the intrinsic losses within the MMI device.  These losses do not enforce a $\varphi=\pi$ phase shift between the probability amplitudes for both photons being reflected and both photons being transmitted. A perfectly balanced 2x2 MMI should operate in exactly the same way as a perfect beamsplitter with the following scattering matrix:

\[
\frac{1}{\sqrt{2}}\left[\begin{array}{cc}
1 & 1\\
1 & -1\end{array}\right]=\frac{1}{\sqrt{2}}\left[\begin{array}{cc}
1 & 1\\
1 & e^{i\pi}\end{array}\right]\]

In the presence of losses, we account for the loss modes by embedding the scattering matrix of the MMI in a larger 4x4 unitary matrix written as:


\[
\left[\begin{array}{cccc}
\left(\eta\overline{\alpha}\right)^{\frac{1}{2}} & \left(\overline{\eta}\overline{\alpha}\right)^{\frac{1}{2}} & \left(\eta\alpha\right)^{\frac{1}{2}} & \left(\overline{\eta}\alpha\right)^{\frac{1}{2}}\\
\left(\overline{\eta}\overline{\alpha}\right)^{\frac{1}{2}} & e^{i\varphi}\left(\eta\overline{\alpha}\right)^{\frac{1}{2}} & e^{i\theta}\left(\overline{\eta}\alpha\right)^{\frac{1}{2}} & e^{i\beta}\left(\eta\alpha\right)^{\frac{1}{2}}\\
\left(\eta\alpha\right)^{\frac{1}{2}} & e^{i\theta}\left(\overline{\eta}\alpha\right)^{\frac{1}{2}} & \cdots & \cdots\\
\left(\overline{\eta}\alpha\right)^{\frac{1}{2}} & e^{i\beta}\left(\eta\alpha\right)^{\frac{1}{2}} & \cdots & \cdots\end{array}\right]\]

where $0\leq\eta\leq1$ is the reflectivity of the MMI, $0\leq\alpha\leq1$ is the loss parameter, $\overline{\eta}=1-\eta$, $\overline{\alpha}=1-\alpha$, $\varphi$ is the internal phase of the MMI, $\theta$ and $\beta$ are the phases of the loss modes. Consequently, the phase between the probability amplitude for both photons being reflected and both photons being transmitted is not necessary $\pi$ anymore. The unitary condition gives a boundary on the internal phase $\varphi$ as a function of the loss $\alpha$:

\[
\left|\cos\left(\frac{\varphi}{2}\right)\right|\leq\frac{\alpha}{1-\alpha}\]

The observed nominal visibility of 88\% could be explained by an intrinsic MMI loss greater than 0.8 dB, which could yield $\varphi=2.74$ instead of $\pi$ phase shift. Indeed, FDTD simulations predict an intrinsic device loss of 0.5 dB, which is within experimental error of the 0.8 dB calculated by the model. By reducing the intrinsic loss of the MMI to 0.2 dB (through improved device design and fabrication), the lower bound on the visibility would improve to 99.5\%. It is worth mentioning that mode propagation simulation shows that the distorsion of the single photon wavepacket ($\sim$100 $\mu$m coherence length in free space) is insignificant in this experiment and does not contribute to a reduction of the visibility. 

\section*{Two-photon entangled state manipulation}

The ability to actively prepare and measure arbitrary quantum states is crucial for the implementation of many quantum information processing experiments. Manipulation of a dual rail-encoded qubits $-$ a single photon in an arbitrary superposition of two optical channels $-$ requires control of the relative phase and amplitude between the two optical paths. Arbitrary single qubit operations can be implemented using just two Hadamard gates (beamsplitter) and phase shifters, with the simple Mach-Zehnder Interferometer (MZI) considered to be most basic fundamental building block required to realise any arbitary N-mode linear quantum photonic circuit \cite{ReckScheme}.

To demonstrate the operation of this fundamental building block, a silicon-on-insulator MZI was formed from two 2x2 MMI couplers and a thermo-optic phase shifter (see Fig.~\ref{Schemes}(b)). The MZI on it's own can apply only a subset of single qubit operations, but any arbitrary single qubit operation can be achieved by simply adding another phase-shifter element before and after the MZI \cite{Matt-NP-3-346}. This device was characterised in both the single-photon and two-photon regime. 

In the first instance, single photons where input into port $a$ (see Fig.~\ref{Schemes}(b) ), and as the internal phase of the MZI was changed, by applying a voltage across the thermo-optic phase shifter, the probability of detecting photons at the output port $c$ varied sinusoidally as $P_c = \frac{1}{2}[1-cos(\phi)]$, with a periodicity of $2\pi$. The observation of this classical interference fringe represents the ability to transform a single-photon input in mode $a$, into a superposition state across modes $c$ and $d$, given by the transformation $\left|10\right\rangle \rightarrow cos(\phi/2) \left|10\right\rangle + sin(\phi/2) \left|01\right\rangle$, where $\phi$ represents the phase shift within the interferometer. In the second instance, the two-photon $\ket{11}$ state was used, with one photon injected into port $a$ and one into port $b$. This input state leads to quantum interference at the first MMI couplers and, after propagating through the phase shifter, transforms to the two-photon entangled state $\frac{1}{\sqrt{2}} ( \left|2\right\rangle \left|0\right\rangle + \exp^{2i\phi} \left|0\right\rangle \left|2\right\rangle )$. Quantum interference at the second MMI coupler enables analysis of this quantum state, giving the probability output of coincidence detections $P_{c-d} = \frac{1}{2}[1-cos(2\phi)]$, with a periodicity of $\pi$ $-$ half that of the single-photon case. 

\begin{figure}
   \includegraphics[width=\columnwidth]{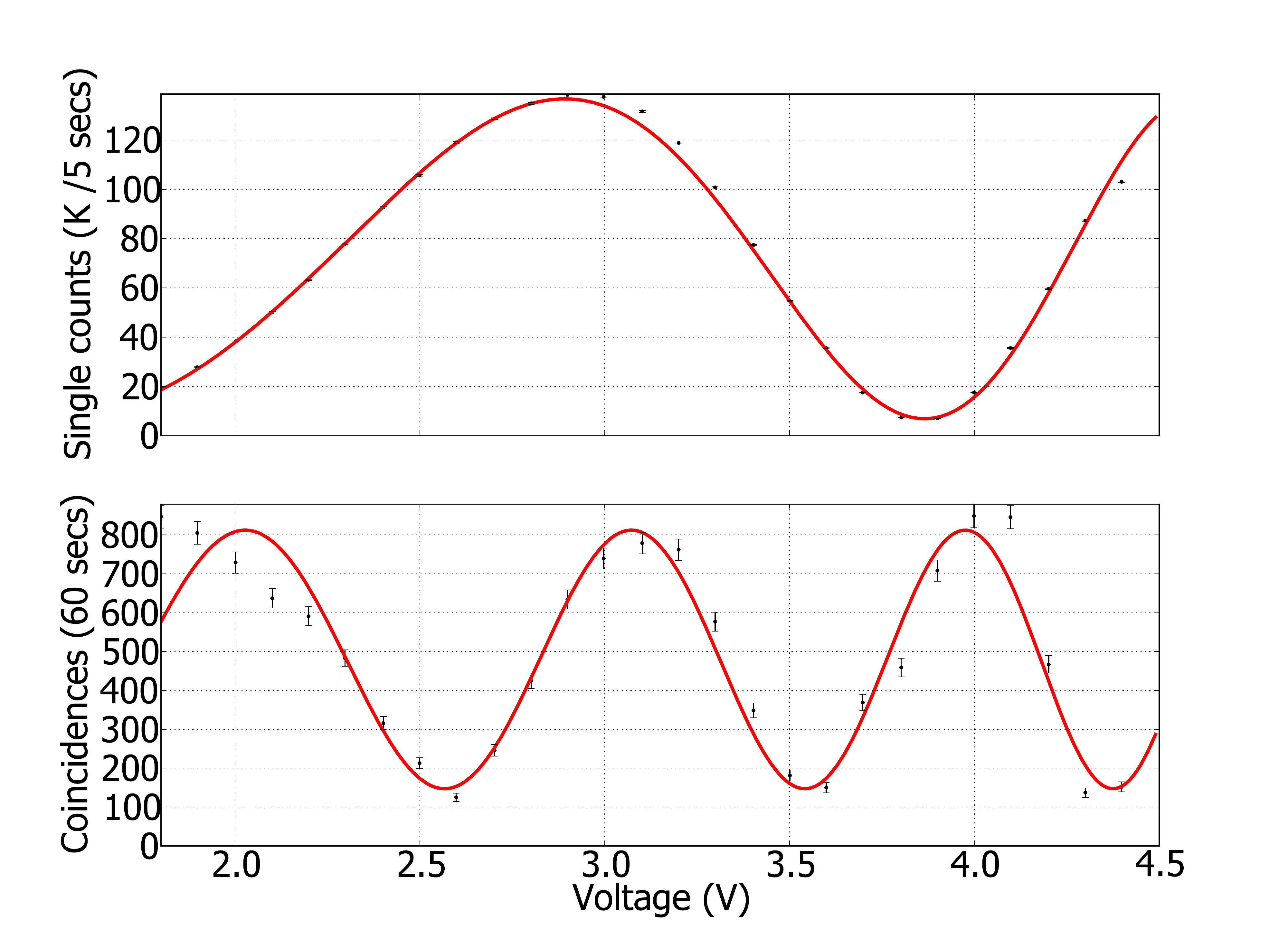}
    \caption{
    \label{fringes}
   \textbf{Single photon and two-photon fringes}
	a) Single photon count rate as a function of voltage applied to the thermo-optic phase-shifter for the input state $\ket{10}$. b) Two-photon coincidence count rate as a function of applied voltage for the input state $\ket{11}$. 
	The $x$ axis is the applied voltage across the thermo-optic phase-shifter. The top figure show the single photon fringe. The $y$ axis represents the number of single photon recorded at one output of the MZI. The bottom figure shows the two-photon fringe. The $y$ axis represents the number of coincidence counts obtained for the two-photon fringe.
	}
\end{figure}

The measured single-photon and two-photon fringes are shown in Fig.~\ref{fringes} for an applied voltage change from 0~V to 4.5~V, resolving 1.5 fringes for the single-photon case, and 3 fringes for the two-photon case. The two-photon fringe had a visibility of $V=81.8\%\pm1.3\%$, which is greater than the threshold $V_{th} = 1/\sqrt{2}$ required to beat the standard quantum limit, demonstrating quantum metrology and the ability to achieve sub shot noise limited measurement.

\section*{Discussion}

This is the first demonstration of interference and manipulation of quantum states of light in silicon integrated quantum circuits, and is a fundamental step for further miniaturisation of photonic quantum circuits. 
All previous integrated waveguide quantum circuits have relied on weakly guided waveguide structures, with a typical index contrast for the silica-on-silicon waveguide technologies of 0.5\%, resulting in a bend radius of greater than 10~mm \cite{Matt-NP-3-346}. This low index contrast and large bend radius result in physically large circuits and components, making future implementations of complex quantum circuit architectures impractical. Due to the high index contrast of the silicon-on-insulator material systems, it is possible to achieve bend radii of below $10\mu m$ -- three orders of magnitude smaller than the silica material system. Individual components can be dramatically reduced in size, such as the 2x2 MMI coupler presented here which is 40 times smaller in length ($27 \mu m$) compared to 1.1~mm length of the equivalent device in silica \cite{ap-nc-2011}. This enables circuits of significantly greater complexity to be realised whilst still maintaining a small chip size, but potentially at the cost of higher waveguide losses. 

The key functions required to realise a fully integrated linear-optic quantum technology are the generation, detection, interference and manipulation of quantum states of light, all on a single optical on-chip. Generation and detection of single photon states in silicon wire waveguide devices has been previously demonstrated \cite{sh-pg-oe-2006,Pernice2011High}, and this work demonstrates the feasibility of on-chip interference and manipulation of quantum states of light in silicon wire waveguide circuits. The high $\chi^3$ nonlinearity of silicon and the high modal confinement of the silicon wire waveguide enables efficient photon pair generation via spontaneous four wave mixing, with a recent demonstration showing indistinguishable photon pair generation from two independent silicon wire waveguides with an external HOM dip visibility of 73\% \cite{NTT-silicon-2011}. High efficiency detection of single-photons has also recently been demonstrated in silicon wire waveguide incorporating superconducting nanowire detectors, with an internal detector efficiency of 94\%  \cite{Pernice2011High}. These recent demonstrations along with the work presented here represent the basic building blocks required to realise an integrated photonic quantum technology platform where quantum states can be generated, manipulated, interfered and detected all on the same circuits, opening up new possibilities in quantum information science and applications.  Ultimately this CMOS compatible technology could be integrated with conventional microelectronic circuits, providing on-chip driver circuits and fast logic.    

\section*{Conclusion}

Quantum interference and manipulation of quantum states of light in silicon wire waveguide circuits has been demonstrated. A maximum HOM dip visibility of $~80\%\pm3\%$ was observed for a 2x2 multimode intereference coupler, and a nominal visibility in the absence of multi-photon terms of $~88\%\pm3\%$ was calculated. Internal MMI losses were shown to be the be dominant mechanism for the reduced nominal visibility. An integrated Mach-Zehnder interferometer with a thermal phase shifter demonstrated on-chip entangled state manipulation with a two photon fringe visibility of  $~81.8\%\pm1.3\%$. These results pave the way for the realisation of fully integrated photonic quantum technologies in silicon.

\section*{Acknowledgement}
This work was supported by NSQI and the European FP7 project QUANTIP. MGT would like to acknowledge the support of the Toshiba Research Fellowship scheme.


\bibliographystyle{unsrt}
\bibliography{bibli_all}



\section{APPENDIX}
This appendix describes the various details of the model used to give the fit in Fig.~\ref{VisibilityvsProbaSPDC}

\subsection{Multiphoton treatment of quantum interference}

The state produced by the SPDC source is assumed to be a non degenerate squeezed state and can be written as:

\[
\left|\Psi\right\rangle =\sqrt{1-\xi^{2}}\sum_{n=0}^{\infty}\xi^{n}\left|n\right\rangle \left|n\right\rangle \]

with $0\leq\xi<1$ where $\xi$ is the squeezing parameter, $n$ is the number of photon pairs generated, $\left|n\right\rangle \left|n\right\rangle$ is the number of photons in the two spatial modes of the source. We label $P_{C1}$,$P_{C2}$ and $P_{CC}$ the probabilities of detecting per pulse, respectively a single photon at detector 1, a single photon at detector 2, and a coincidental event between the two detectors C1 and C2.

Those probabilities can be expressed as:

\begin{eqnarray*}
P_{C1} & = & \sqrt{(1-\xi^{2})}\sum_{n=1}^{\infty}\xi^{2n}\left(1-\left(1-\eta_{1}\right)^{n}\right)\\
P_{C2} & = & \sqrt{(1-\xi^{2})}\sum_{n=1}^{\infty}\xi^{2n}\left(1-\left(1-\eta_{2}\right)^{n}\right)\\
P_{CC} & = & \sqrt{(1-\xi^{2})}\sum_{n=1}^{\infty}\xi^{2n}\left(1-\left(1-\eta_{1}\right)^{n}\right)\left(1-\left(1-\eta_{2}\right)^{n}\right)\end{eqnarray*}

with $\eta_{1}$ and $\eta_{2}$ being the overall efficiencies (including collection, transmission and detection) of respectively channels 1 and 2.

We measured the quantities C1 (number of detection events per second at the detector 1), C2 (number of detection events per second at the detector 2) and CC (number of coincidental detection events per second at the detector 1 and 2) for different pump powers.

From those datasets, we ran a minimisation methods to relate the different count rates $C1$, $C2$, $CC$ obtained from the pump powers $I_{k}$ to the channel efficiencies $\eta_{1}$ and $\eta_{2}$, and the $\xi^{2}(I_{k})$ parameter using the following equations:

\begin{eqnarray*}
C1(I) & = & f\times P_{C1}(I)\\
C2(I) & = & f\times P_{C2}(I)\\
CC(I) & = & f\times P_{CC}(I)\end{eqnarray*}

where $f=80$ MHz is the repetition rate of the pulsed laser and $I$ is the intensity of the pump.

We then launched the state $\left|\Psi\right\rangle $ for different pump powers into the MMI coupler, and recorded quantum interference patterns with visibilities $V(I)$. In order to infer the visibility at low intensity, we applied the model described below.

\subsection{Theoretical visibility}

\begin{figure}
  \includegraphics[width=\columnwidth]{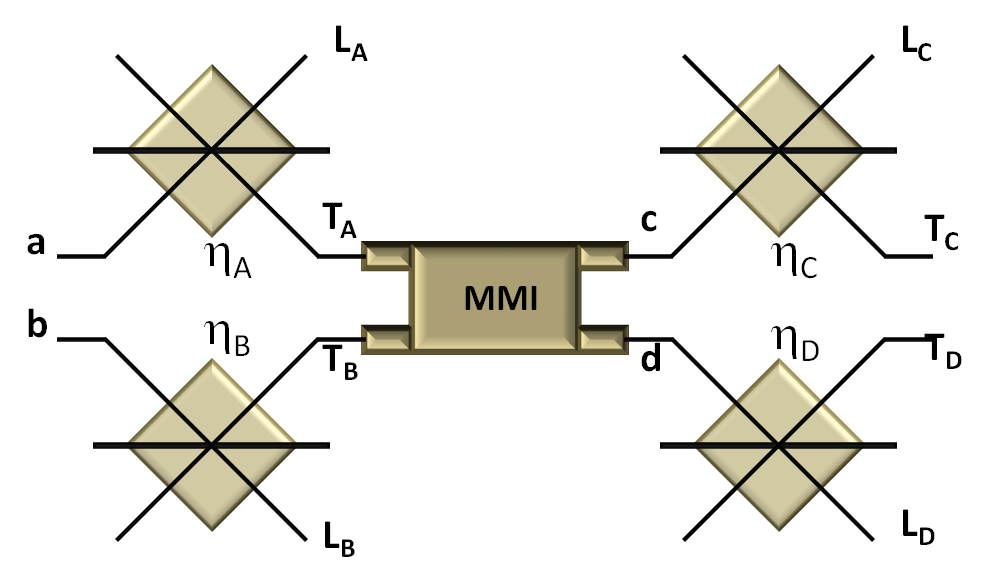}
  \caption{Input and output losses modeling of a 2x2 MMI coupler\label{fig:MMIloss}}

\end{figure}

The theoretical visibility is expressed as $1-\frac{P_{I}}{P_{D}}$ where $P_{I}$ is the probability to get a coincidental event when the photons generated in the two modes are indistinguishable, and $P_{D}$ is the probability to get a coincidental event when the two photons are distinguishable. If we used a true single photon source in an ideal lossless circuit, $P_{I}=0$, $P_{D}=0.5$ and $V=1$. However, we used a squeezed state $\left|\Psi\right\rangle $ as the input state, and the setup exhibits collection losses, propagation losses and non unitary detection efficiencies. We modelled those losses by adding a virtual beamsplitter at each input and output port of the MMI coupler with reflectivity's $\eta_{A}$,$\eta_{B}$ and $\eta_{C}$,$\eta_{D}$ as shown in Fig.~\ref{fig:MMIloss}. And we trace over the loss modes $L_{A},L_{B},L_{C},L_{D}$. All that is needed is to compute $P_{I}$ and $P_{D}$ taking into account these losses.

We start with a pure squeezed state with a density matrix $\hat{\rho}=\left|\Psi\right\rangle \left\langle \Psi\right|$, we first propagate the state through the two input beamsplitters, we then trace over the loss modes $L_{A}$and $L_{B}$. We further propagate the obtained reduced density matrix through the MMI coupler, assuming a 50/50 splitting ratio. And we finally propagate the output state through the last two loss beam splitter and trace over the modes $L_{C}$ and $L_{D}$.

By writing $\hat{\rho}_{I\! out}$ as the final density matrix, we compute the probability to get a coincidence by calculating:

\[
P_{I}=\sum_{n=1}^{\infty}\sum_{m=1}^{\infty}\left\langle n\right|\left\langle m\right|\hat{\rho}_{I\! out}\left|n\right\rangle \left|m\right\rangle \]

We implemented an algorithm to compute this probability to an arbitrary precision and found that computing the terms up to $n,m=9$ was accurate for the different pump power we used. Adding higher order terms did not changed significantly the results.


For the case where the photons are distinguishable, we start with $\left|\Psi_{dist}\right\rangle$ defined below and assume that the photons are orthogonal. In practice,
one arm is delayed with respect to the other and we consider the photons launched in mode $a$ at time $t_{1}$ and the photons launched in mode $b$ at time $t_{2}$, giving:

\[
\left|\Psi_{dist}\right\rangle =\sqrt{1-\xi^{2}}\sum_{n=0}^{\infty}\xi^{n}\left|n\right\rangle_{t_1} \left|n\right\rangle_{t_2} \]

Then applying the same evolution as previously for the indistinguishable photon case, we compute the probability $P_{D}$ of getting a coincidence when the photons are distinguishable.

\subsection{Fit of the experimental data}

We recorded the visibility as a function of the pump power. For each intensity, we compute the associated squeezing parameter ($\xi$ ). The collection and detector efficiencies and the device losses are known. The only free parameter over which we need to minimize is the visibility. We therefore use an equation of the form:

\[V\left(\xi\right)=1-\frac{\alpha P_{I}\left(\xi\right)+\left(1-\alpha\right)P_{D}\left(\xi\right)}{P_{D}\left(\xi\right)}\]

where $\alpha$ is the parameter which quantifies the overlap between the photons which includes residual spectral entanglement from the source and imperfections in the MMI coupler. Using this model, we extrapolate the nominal visibility that would be expected at very low pump power.

\end{document}